\documentclass{appolb}
\usepackage{epsfig}
\newcommand{\eq}{\begin{equation}}
\newcommand{\eqx}{\end{equation}}
\newcommand{\eqn}{\begin{eqnarray}}
\newcommand{\eqnx}{\end{eqnarray}}
\def\lapproxeq{\lower .7ex\hbox{$\;\stackrel{\textstyle
<}{\sim}\;$}}
\def\gapproxeq{\lower .7ex\hbox{$\;\stackrel{\textstyle
>}{\sim}\;$}}
\begin{document}
\title{GVMD model predictions for the low $Q^2$ behaviour 
of the spin 
structure function $g_1(x,Q^2)$ and of the DHGHY integral $I(Q^2)$%
\thanks{Presented at 10th International Workshop 
DIS2002, Cracow, 30 April - 4 May 2002. }%
}
\author{\underline{Barbara~Bade\l{}ek} $^{a,\,b}$, Jan~Kwieci\'nski $^c$ 
$\rm and$ Beata~Ziaja $^{d,\,e}$
\address{$^a$ Institute of Experimental Physics, Warsaw University,\\
                       Ho\.za 69, 00-681 Warsaw, Poland  \\
         $^b$ Department of Physics, Uppsala University, \\
                       P.O. Box 530, S-751 21 Uppsala, Sweden \\
         $^c$ Department of Theoretical Physics,
                       H. Niewodnicza\'nski Institute of Nuclear Physics, 
                       Radzikowskiego 152,  31-342 Cracow, Poland \\
         $^d$ Department of Biochemistry, BMC, Uppsala University, \\
                       Box 576, S-751 23 Uppsala, Sweden \\
         $^e$ High Energy Physics, Uppsala University, \\
                       P.O. Box 535, S-751 21 Uppsala, Sweden}\\
}

\maketitle
\begin{abstract}
Predictions for $g_1(x,Q^2)$ at low $Q^2$ are obtained in the framework of the
GVMD model. Contributions from both light and heavy vector mesons are evaluated.
The DHGHY sum rule  is employed to fix the magnitude of the light vector meson
contribution to $g_1$, using the recent measurements in the region
of baryonic resonances. The DHGHY moment function is calculated.
Predictions are compared to the data. 
\end{abstract}
  
\section{Introduction}
\noindent
Data on polarized nucleon structure function $g_1(x,Q^2)$
are now available at 
low values of (negative) four-momentum transfer,
$Q^2$, \cite{nne143smc98hermes,ssmc98}.
This is of particular interest since
nonperturbative mechanisms dominate the particle dynamics there and 
a transition from soft- to hard physics may be studied.
\\

\noindent
In the previous attempt, \cite{bkk}, $g_1$ at low $x$ and low $Q^2$ was 
described within a formalism based on the unintegrated spin dependent parton
distributions,
incorporating the leading order Altarelli--Parisi evolution and the double
ln$^2$(1/$x$) resummation at low $x$. A VMD-type nonperturbative
part of $g_1$ was also included, its unknown normalisation was  extracted 
from the data and turned out to be nonzero and negative.  \\

\noindent
In this paper we apply the Generalized Vector Meson Dominance (GVMD) model.
to evaluate the nonperturbative contributions
to the polarized structure function $g_1(x,Q^2)$ at low values of $Q^2$.
The heavy meson ($M_V>Q_0$) contribution is
directly related to the structure function in the scaling region, $g_1^{AS}$,
described by the QCD improved parton model, suitably extrapolated to the
low $Q^2$ region. The contribution
of light ($M_V<Q_0$) vector mesons describes nonperturbative effects and
 vanishes as 1/$Q^4$ for large $Q^2$. At low $Q^2$ these effects are large
and predominant. Here  $M_V$ denotes the mass of a vector meson.
Then the
Drell-Hearn-Gerasimov-Hosoda-Yamamoto (DHGHY) sum rule \cite{DH}
together with measurements in the resonance region are employed
to fix the magnitude of the light vector meson contribution
to $g_1$. 

\section{The GVMD representation of the structure fun\-ction $g_1(x,Q^2)$
and the DHGHY sum rule}
\noindent
In the GVMD model,  $g_1$
has the following representation, valid for fixed $W^2\gg Q^2$,
i.e. small values of $x$, $x=Q^2/(Q^2+W^2-M^2)$:
{\footnotesize
\eq
g_1(x,Q^2)=g_1^{L}(x,Q^2)+g_1^{H}(x,Q^2)=\frac{M\nu}{4\pi}\,\sum_V\,
\frac{M^4_V \Delta \sigma_V(W^2)}{\gamma_V^2(Q^2+M^2_V)^2}+
g_1^{AS}({\bar x},Q^{2}+Q_0^2).
\label{gvmd}
\eqx
}
\noindent
The first term
%
%
sums up contributions from light vector mesons, $M_V < Q_0$ where  $Q_0^2
\sim$ 1 GeV$^2$ \cite{el89}.
 Here $W$ is the invariant mass of the electroproduced
hadronic system, $\nu=Q^2/2Mx$, and $M$ is the nucleon mass.
The constants $\gamma_V^2$ are determined from the leptonic widths of the
vector mesons and the cross sections $\Delta \sigma_V(W^2)$ are combinations
of the total cross sections for the scattering of polarised mesons and nucleons.
They are not known and have to be
parametrized. Following Ref.\ \cite{bkk}, we assume that they can be expressed
through the combinations of nonperturbative parton distributions,
$\Delta p_j^{(0)}(x)$, evaluated at fixed $Q_0^2$.
{\footnotesize
}
The second term in (\ref{gvmd}), $g_1^{H}(x,Q^2)$, which represents
the contribution of heavy ($M_V > Q_0$) vector mesons to $g_1(x,Q^2)$
can also be treated as an extrapolation of the QCD improved parton model
structure function, $g_1^{AS}(x,Q^{2})$, to arbitrary values of $Q^2$. Here
%
%
the scaling variable $x$ 
is replaced by ${\bar x}=(Q^2+Q_0^2)/(Q^2+Q_0^2+W^2-M^2)$, \cite{el89}.
It follows  that
 $g_1^{H}(x,Q^2)\rightarrow g_1^{AS}(x,Q^{2})$ as $Q^2$ is large.
We thus get:
{\footnotesize
\eqn
g_1(x,Q^2)&=&
%
C\left[ \frac{4}{9}(\Delta u_{val}^{(0)}(x)+\Delta \bar u^{(0)}(x))
+\frac{1}{9}(\Delta d_{val}^{(0)}(x)+\Delta \bar d^{(0)}(x))\right]
\frac{M^4_{\rho}}{(Q^2+M^2_{\rho})^2}\nonumber\\
&+&C\left[ \frac{1}{9}(2\Delta \bar s^{(0)}(x))\right]\frac{M^4_{\phi}}
{(Q^2+M^2_{\phi})^2}\nonumber \\
&+&g_1^{AS}({\bar x},Q^{2}+Q_0^2).\label{gvmdu}
\eqnx
}

\noindent
The only free parameter in (\ref{gvmdu}) is the constant $C$. Its value
may be fixed in the photoproduction limit where the first moment of
$g_1(x,Q^2)$ is related to the anomalous magnetic moment of the
nucleon via the DHGHY sum rule, cf.\ \cite{ioffe,ioffe2}:
\eq
I(0)=I_{res}(0)
+ M\,\int_{\nu_t(0)}^{\infty}\,\frac{d\nu}{\nu^2}\,g_1\left(x(\nu),0\right)
=-\kappa^2_{p(n)}/4.
\label{i0}
\eqx
where the DHGHY moment
%
before taking the $Q^2$=0 limit has been split into two parts, corresponding
to $W < W_t\sim$ 2 GeV (baryonic resonances) and $W > W_t$:
\eq
I(Q^2)=I_{res}(Q^2)
+ M\,\int_{\nu_t(Q^2)}^{\infty}\,\frac{d\nu}{\nu^2}\,g_1\left(x(\nu),Q^2\right),
\label{iq2}
\eqx
Here $\nu_t(Q^2)=(W_t^2+Q^2-M^2)/2M$. Substituting $g_1\left(x(\nu),0\right)$ 
in Eq. (\ref{i0}) by Eq. (\ref{gvmdu}) at $Q^2=0$ 
we may obtain the value of $C$ from (\ref{i0}) if $I_{res}(0)$, the contribution
from resonances, is known e.g. from measurements.


\section{Numerical calculations for the proton}
\noindent
To obtain the value of $C$ from Eq.
(\ref{i0}), $I_{res}(0)$ was evaluated using the preliminary
 data taken at ELSA/MAMI by the GDH Collaboration \cite{elsa}
at the photoproduction, for $W_t$=1.8 GeV. The $g_1^{AS}$  
was parametrized using GRSV2000 fit \cite{grsv2000} for the ``standard 
scenario'' at the NLO accuracy. The $\Delta p_j^{(0)}(x)$ in Eq.(\ref{gvmdu}) 
were evaluated at fixed $Q^2 = Q_0^2$, using, either (i) the GRSV2000 fit,
or (ii) a simple, ``flat'' input, $\Delta p_i^{(0)}(x)=N_i (1-x)^{\eta_i}$
with $\eta_{u_v}=\eta_{d_v}=3,$  $\eta_{\bar u} = \eta_{\bar s} = 7 $
and $\eta_g=5$, \cite{STRATMAN}.
We have assumed $Q_0^2$ = 1.2 GeV$^2$ 
as in the analysis of $F_2$, \cite{el89}. 
As a result the constant $C$ was found to be --0.30 in case (i) and
--0.24 in case (ii). These values change at most by 13$\%$ when
$Q_0^2$ changes in the interval 1.0$ < Q_0^2 < $1.6 GeV$^2$.
\noindent
Negative value of the nonperturbative, Vector Meson Dominance,
contribution was also obtained in
\cite{bkk} and from the phenomenological analysis
of the sum rules \cite{ioffe2,burkert}. \\

\begin{figure}[t]
\begin{center}
\setlength{\unitlength}{0.1mm}
\begin{picture}(1200,650)
\put( -350, 0){
\epsfig{width=6cm, height=5.5cm, file=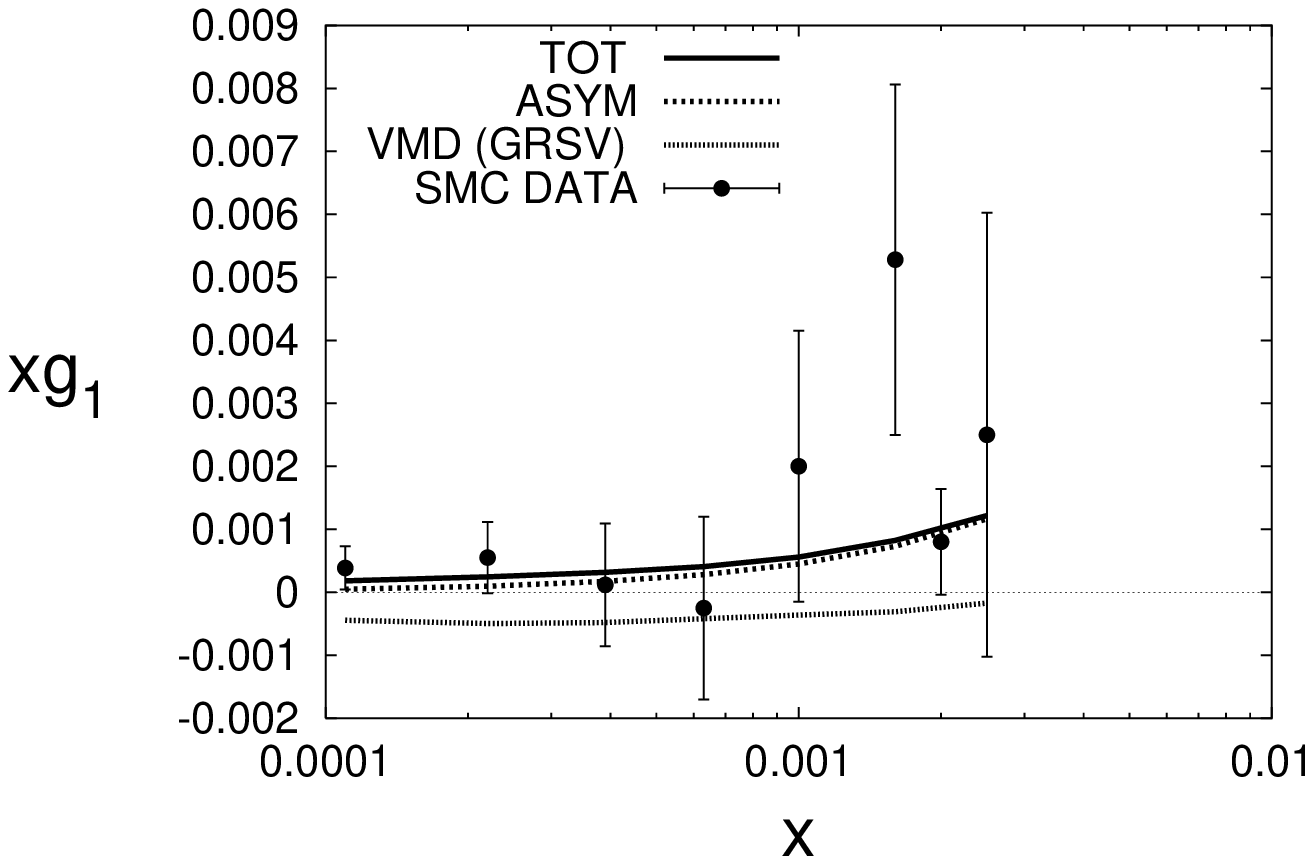}
}
\put(300, 0){
\epsfig{width=6cm, height=5.5cm, file=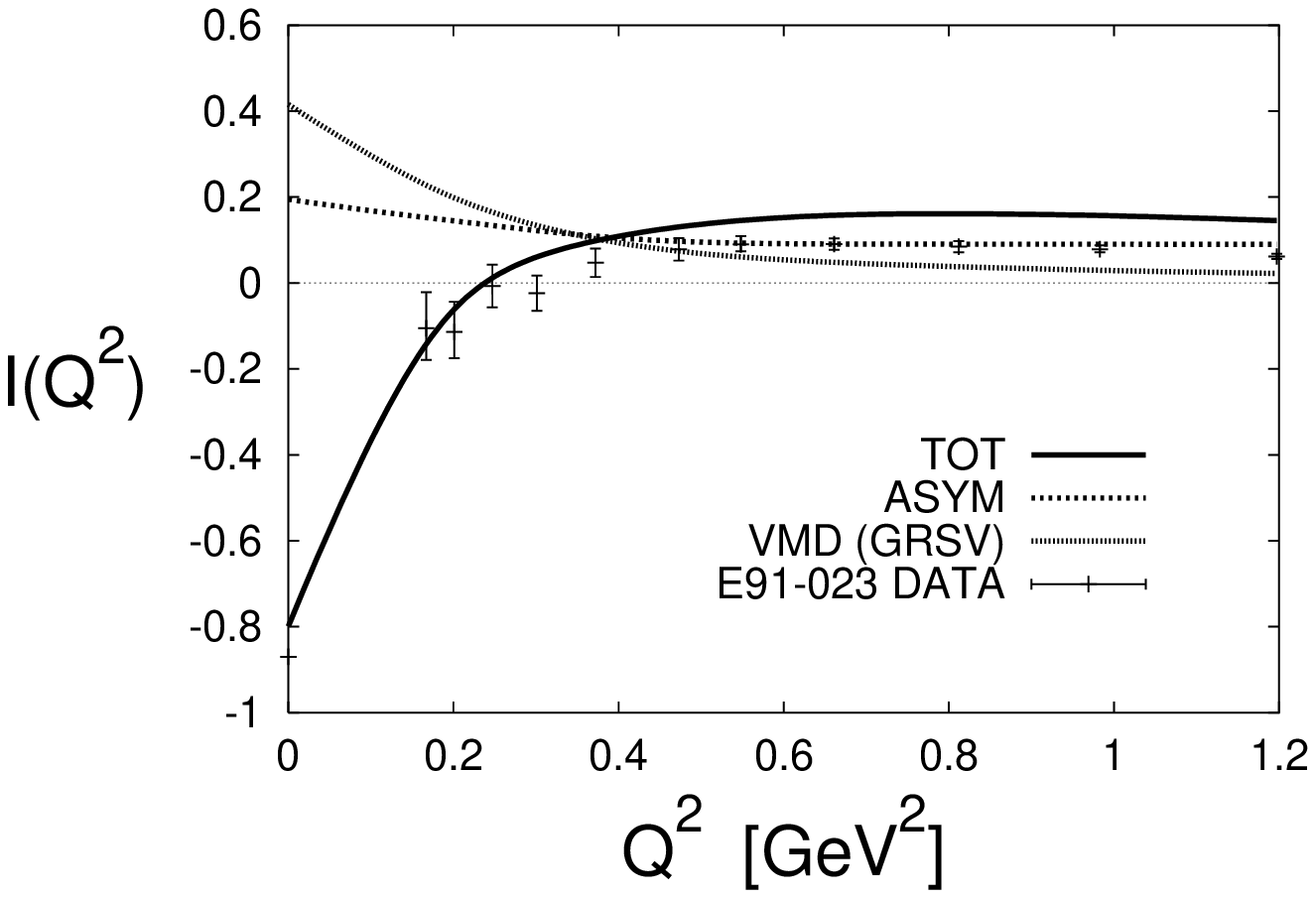}
}
\end{picture}
\end{center}
\caption{a) Values of $xg_1$ for the proton as a function of $x$ at the
measured values of $Q^2$ in the non-resonant region, $x<x_t=Q^2/2M\nu_t(Q^2)$.
Both the VMD input and $g_1^{AS}$ have been evaluated using the GRSV 
fit for standard scenario at the NLO accuracy \cite{grsv2000}. 
Contributions of the VMD and of the $xg_1^{AS}$
are shown separately. Points are the SMC measurements at
$Q^2 <$ 1 GeV$^2$, \cite{ssmc98}; errors are total. The
curves have been calculated at the measured $x$ and $Q^2$ values. 
b) The DHGHY moment $I(Q^2)$ for the proton. Details as in Fig.1a.
Points mark the contribution of resonances as measured by the JLAB E91-023,
\cite{e91-023} at $W < W_t(Q^2)$.
}
\label{fig1}
\end{figure}

\noindent
Our $g_1$, Fig.\ref{fig1}a, reproduces well
a general trend in the data; however experimental errors are too large for
a more detailed analysis. To compute the DHGHY moment, Eq.(\ref{iq2}), 
for the proton, we used the preliminary results of the JLAB E91-023 
experiment \cite{e91-023} for 0.15$\lapproxeq Q^2 \lapproxeq $1.2 GeV$^2$ 
and $W < W_t=W_t(Q^2)$~\cite{fatemi}. Results, Fig.\ref{fig1}b, show that
partons contribute significantly even at $Q^2\rightarrow$ 0 where the
main part of the $I(Q^2)$ comes from resonances. \\
%
%
%
\noindent
\noindent
%
%
\noindent
\begin{figure}[t]
\begin{center}
\epsfig{width=8cm, height=5.3cm, file=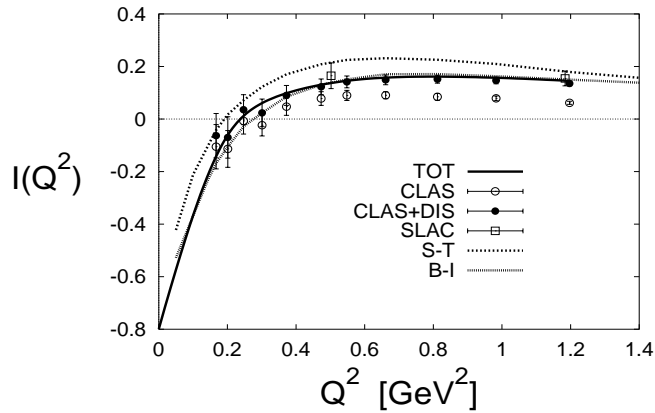}\\
\end{center}
\caption{The DHGHY moment $I(Q^2)$ for the proton with the VMD part parametrized
using the GRSV fit \cite{grsv2000}.
Shown are also calculations of \cite{burkert}
(``B--I'') and \cite{soffer} (``S--T''). Points marked ``CLAS'' are from
the JLAB E91-023 experiment 
\cite{e91-023}: the open circles refer to the resonance region, $W < W_t(Q^2)$
and the full circles contain a correction for the DIS contribution. 
Errors are total.
}
\label{fig3}
\end{figure}
%
%

\noindent
In Fig. \ref{fig3} we show our DHGHY moment together with the results 
of calculations of Refs \cite{burkert,soffer} as well as with
the E91-023 measurements in the resonance region used as an input to our
$I(Q^2)$ calculations. We also show the E91-023 data
corrected by their authors for the deep inelastic contribution.
Our calculations are slightly
larger than the DIS-corrected data and then the results of \cite{burkert} but
clearly lower than the results of \cite{soffer} which overshoot the data.\\

\noindent
It is a pleasure to thank to the organizers for the splendid workshop. \\

\noindent
This research has been supported in part by the Polish Committee for Scientific  Research with grants 2 P03B 05119, 2PO3B 14420 and European
Community grant 'Training and Mobility of Researchers', Network 'Quantum
Chromodynamics and the Deep Structure of Elementary Particles'
FMRX-CT98-0194. 
B.\ Z.\ was supported by the Wenner-Gren Foundations.

\end{document}